\documentclass[12pt,oneside,a4paper]{article}

\usepackage{hyperref}
\hypersetup{colorlinks,bookmarksopen,bookmarksnumbered,citecolor=blue,
linkcolor=black,pdfstartview=FitH,urlcolor=blue}

\usepackage[dvipdfmx]{graphicx}
\usepackage[hang,small,bf]{caption}
\usepackage[subrefformat=parens]{subcaption}
\usepackage{amssymb}
\usepackage{cite}
\usepackage{bm}
\usepackage{indentfirst}
\usepackage{amsmath}
\usepackage{hhline}
\usepackage{multirow}
\usepackage{color}
\usepackage{adjustbox}

\usepackage{braket}
\usepackage{textcomp}

\allowdisplaybreaks
\numberwithin{equation}{section}

\def\nn{\nonumber}

\definecolor{dred}{rgb}{0.7,0.15,0.09}
\definecolor{dblue}{rgb}{0,0.12,0.64}
\definecolor{dgreen}{rgb}{0.2,0.51,0.19}

\begin{document}

\begin{titlepage}


\begin{center}

\vspace*{2cm}
{\LARGE\textbf{
Analysis of inflationary models \\
in higher-dimensional uniform inflation
}
 }
\vspace{1cm}

\renewcommand{\thefootnote}{\fnsymbol{footnote}}
{\Large Takuya Hirose}\footnote[1]{t.hirose@ip.kyusan-u.ac.jp}
\vspace{5mm}

\textit{
 {Faculty of Science and Engineering, Kyushu Sangyo University, \\
 Fukuoka 813-8503, Japan}\\
}

\vspace{10mm}

\abstract{
\noindent
We consider higher-dimensional uniform inflation, in which the extra dimensions expand at the same rate as three-dimensional non-compact space during inflation.
We compute the cosmological perturbation in $D+4$ dimensions and derive the spectral index $n_s$ and the tensor-scalar ratio $r$.
We analyze five inflationary models: chaotic inflation, natural inflation, quartic hilltop inflation, inflation with spontaneously broken SUSY, and $R^2$ inflation.
By combining the results from these models with the Planck 2018 constraints, we discuss that it is not desirable for the extra-dimensional space to expand at the same rate as the three-dimensional non-compact space, except for the case of one extra dimension.
}

\end{center}
\end{titlepage}

\renewcommand{\thefootnote}{\arabic{footnote}}
\newcommand{\bhline}[1]{\noalign{\hrule height #1}}
\newcommand{\bvline}[1]{\vrule width #1}

\setcounter{footnote}{0}

\setcounter{page}{1}

\section{Introduction}
Inflation \cite{Guth:1980zm} provides a solution to the problems Big Bang faces: mainly, the horizon problem and the flatness problem (for a review, see \cite{Mukhanov:1990me, Riotto:2002yw, Baumann:2009ds, Langlois:2010xc}).
The most interesting inflationary model is slow-roll inflation.
In this model, a scalar field called inflaton is introduced.
The potential of inflaton must be flat, and it is characterized by slow-roll parameters.
Although there are many models of slow-roll inflation \cite{Martin:2013tda}, they must satisfy the Planck 2018 constraints \cite{Planck:2018vyg, Planck:2018jri}.

Recently, ``higher-dimensional uniform inflation'' \cite{Anchordoqui:2022svl, Anchordoqui:2023etp} was proposed.
The idea is that compact extra dimensions can acquire large size by higher-dimensional uniform inflation.
This inflationary scenario is motivated by the cosmological hierarchy problem (the hierarchy between particle physics and cosmology), rather than the electroweak hierarchy problem \cite{Arkani-Hamed:1998jmv, Antoniadis:1998ig, Arkani-Hamed:1998sfv}\footnote{Inflation with large extra dimensions is studied in \cite{Arkani-Hamed:1999fet, Cline:1999ky, Nihei:1999mt}.}.
The cosmological hierarchy problem is related to the dark dimension \cite{Montero:2022prj}, based on the distance conjecture or swampland program \cite{Lust:2019zwm} and several experimental bounds.
In the dark dimension proposal, the size of extra dimensions is predicted in the range of 0.1\,$\mu$m to 10\,$\mu$m.
On the other hand, the power spectrum of primordial density fluctuations is consistent with cosmic microwave background (CMB) observations for 1\,$\mu$m$\sim$10\,$\mu$m size of one extra dimension \cite{Anchordoqui:2023etp, Anchordoqui:2024amx}.
These results imply that the higher-dimensional uniform inflation is strongly related to the dark dimension.



In the higher-dimensional uniform inflation scenario, the extra dimensions expand at the same rate as three-dimensional non-compact space during inflation.
Under this assumption,
the cosmological perturbations from five dimensions are studied in \cite{Antoniadis:2023sya}.
In \cite{Antoniadis:2023sya}, it is pointed out that the spectral index $n_s$ and tensor-scalar ratio $r$ from five dimensions are modified from the standard four-dimensional case.
Since $n_s$ and $r$ are constrained by $n_s$-$r$ contour in Planck 2018 results \cite{Planck:2018jri},
we can determine whether an inflationary model in higher dimensions is allowed or excluded if we specify the inflaton potential in higher dimensions.

In this paper, we consider the $(D+4)$-dimensional uniform inflation, where $D$ is the number of extra dimensions.
Following \cite{Anchordoqui:2023etp}, we require that the compact extra dimensions expand at the same rate as three-dimensional non-compact space during inflation.
We compute the cosmological perturbation in $D+4$ dimensions and derive the spectral index $n_s$ and the tensor-scalar ratio $r$.
We analyze five inflationary models motivated by \cite{Planck:2018jri}: chaotic inflation, natural inflation, quartic hilltop inflation, inflation with spontaneously broken SUSY, and $R^2$ inflation.
We also compare our results with Planck 2018 data.

This paper is organized as follows.
In section \ref{setup}, we introduce our setup and define the inflationary parameters in $D+4$ dimensions.
We compute the cosmological perturbation in $D+4$ dimensions in section \ref{CosmoPer}.
At the end of this section, the spectral index and tensor-scalar ratio in $D+4$ dimensions are summarized.
In section \ref{ex}, we analyze the five inflationary models and compare the results with $n_s$-$r$ contour.
We provide our conclusion in section \ref{discuss}.
Analysis of orbifold and the derivation of $R^2$ inflation are summarized in appendices.

\section{Set up}\label{setup}
\subsection{Background metric and Friedmann equations}
We consider $(D+4)$-dimensional spacetime.
The $D$-dimensional extra space is compactified as $S^1\times S^1\times\cdots S^1$ with the same radius.
The $(D+4)$-dimensional background metric is given by
\begin{align}
ds^2&=-dt^2+a(t)^2d\vec{x}^2_3+b(t)^2d\vec{y}^2_D,
\label{BGmetric}
\end{align}
where $\vec{x}_3$ is the coordinate in three-dimensional directions and $\vec{y}_D$ is the coordinate in extra-dimensional directions.
$a(t)$ and $b(t)$ are the scale factors in three and extra dimensions, respectively.
We are interested in the case that the expansion coefficients of $a(t)$ and $b(t)$ are the same.
This situation is represented by
\begin{align}
a(t)=a_0 e^{Ht},\quad
b(t)=b_0e^{Ht}\quad\Rightarrow\quad
H=\frac{\dot{a}}{a}=\frac{\dot{b}}{b},
\label{scalefactor}
\end{align}
where $H$ is Hubble parameter.
Hereafter, we take $a_0=1$ for simplicity.
$b_0$ is an initial radius of the extra dimensions.
Defining the conformal time $\tau=e^{-Ht}/H$,  the metic \eqref{BGmetric} has
\begin{align}
ds^2&=a^2(\tau)\Big(-dt^2+d\vec{x}^2_3+b^2_0 d\vec{y}^2_D\Big).
\label{BGmetric:conformal}
\end{align}

From the metric \eqref{BGmetric:conformal}, we can derive Einstein equations or Friedmann equations with the extra dimensions \cite{Arkani-Hamed:1999fet, Cline:1999ky, Nihei:1999mt, Anchordoqui:2023etp}.
The results are given by
\begin{align}
&
\frac{(D+2)(D+3)}{2}\mathcal{H}^2=\frac{a^2(\tau)\rho}{M^2_{pl}}, \label{FD:ACC} \\
&
(D+2)\left(\mathcal{H}'+\frac{D+1}{2}\mathcal{H}^2\right)=-\frac{a^2(\tau)p}{M^2_{pl}}, \label{FDp:ACC}
\end{align}
where $\mathcal{H}$ means the conformal Hubble parameter and the prime represents the derivative of the conformal time $\tau$.
$M_{pl}=2.4\times10^{18}$ GeV is the four-dimensional reduced Planck mass.
Introducing the Planck mass in $D+4$ dimensions $M_*$, $M_{pl}$ is related to $M_*$ as 
\begin{align}
M^2_{pl}=b^D M^{D+2}_{*}.
\label{Planckmass}
\end{align}

$\rho$ and $p$ are the energy density and the pressure, respectively.
Note that the pressure in the extra-dimensional directions (denoted $p_D$) is equivalent to the pressure in three dimensions (denoted $p_3$) since the Hubble rate is the same for $a(t)$ and $b(t)$.
Therefore, we denote $p=p_3=p_D$.
Combining Eq. \eqref{FD:ACC} with Eq. \eqref{FDp:ACC}, the useful formula is obtained:
\begin{align}
(D+2)(\mathcal{H}^2-\mathcal{H}')=\frac{a^2}{M^2_{pl}}(\rho+p). \label{FDpr:ACC}
\end{align}
From the conservation law of energy, we obtain
\begin{align}
\rho'+(D+3)\mathcal{H}(\rho+p)=0.
\label{CoE}
\end{align}

\subsection{Inflationary parameter in $D+4$ dimensions}
We introduce a single inflaton $\phi$ with a potential $V(\phi)$ in $D+4$ dimensions.
\footnote{The inflaton $\phi$ has $(D+2)/2$ mass dimensions in $D+4$ dimensions.
In this paper,  we normalize the inflaton to one mass dimension using the radius of the extra dimensions.
}
In this section, we discuss arbitrary inflaton potentials, not specify the inflationary models.
We will discuss the specific models of inflation in a later section.

Introducing the inflaton, the energy density and the pressure are represented by
\begin{align}
\rho=\frac{(\phi')^2}{2a^2}+V,\quad
p=\frac{(\phi')^2}{2a^2}-V.
\label{EandP}
\end{align}
Substituting Eq. \eqref{EandP} for Eq. \eqref{CoE}, we get the equation of motion for inflaton as
\begin{align}
\phi''+(D+2)\mathcal{H}\phi'+a^2(\tau)V_{\phi}=0,
\label{EoM:ACC}
\end{align}
where $V_\phi=dV/d\phi$.
In the slow-roll conditions $(\dot{\phi})^2\ll V(\phi)$, we introduce two slow-roll parameters as
\begin{align}
\epsilon=1-\frac{\mathcal{H}'}{\mathcal{H}},\quad
\delta=1-\frac{\phi''}{\mathcal{H}\phi'}.
\end{align}
Using Eqs. \eqref{FD:ACC} and \eqref{EoM:ACC} with $(\dot{\phi})^2\ll V(\phi)$, we define the two potential slow-roll parameters as
\begin{align}
\epsilon_V=\frac{D+2}{4}M^2_{pl}\left(\frac{V_\phi}{V}\right)^2,\quad
\eta_V=\frac{D+2}{2}M^2_{pl}\frac{V_{\phi\phi}}{V}.
\label{SRparameters}
\end{align}
In the slow-roll regime, $\epsilon\simeq\epsilon_V$ and $\delta=\eta_V-\epsilon_V$ are satisfied.
Note that inflation ends when the slow-roll conditions are violated as $\epsilon_V(\phi_{\text{end}})\approx1$.

The number of e-folds $N_*$ is defined by
\begin{align}
N_*\equiv\ln\frac{a_{\text{end}}}{a}=\int_{t}^{t_{\text{end}}}Hdt=\frac{2}{(D+2)M^2_{pl}}\int_{\phi_{\text{end}}}^{\phi_*}\frac{V}{V_\phi}d\phi,
\label{e-folds}
\end{align}
where $\phi_*$ corresponds to the field value of the CMB observation.
According to CMB data, one finds $40\le N_*\le60$.
We note that the precise value of $N_*$ is estimated by the detail of reheating after inflation.
From Eqs. \eqref{scalefactor} and \eqref{e-folds}, we obtain $b_{\text{end}}=b_0e^{N_*}$.
Combining Eq.\eqref{Planckmass} with $b_{\text{end}}$, $M_*$ is expressed as
\begin{align}
M_*=(b_0 e^{N_*})^{-D/(D+2)}M^{2/(D+2)}_{pl}.
\label{e-folds_M*}
\end{align}

Let us take $b_{\text{end}}\sim10\,\mu$m for $D=1$ and $N_*=60$, motivated by the dark dimension \cite{Montero:2022prj} as an example. the initial radius is estimated by $b_0\sim10^{-25}\,\mu$m.
The scale of $b^{-1}_0$ roughly becomes $b^{-1}_0\sim10^{15}$\,GeV.

We comment on the end of inflation.
After the inflation, the radion $b$ must be stabilized, while three-dimensional non-compact space follows the four-dimensional evolution.
We assume that the radion stabilization could be realized since it is beyond the scope of this paper.
The details of radion stabilization are discussed in \cite{Antoniadis:2023sya}.

\subsection{Size of extra dimension and experimental constraint}
The size of extra dimensions is bounded by the torsion balance experiments \cite{Lee:2020zjt} for $D=1$, the constraint of the neutron star excess heat \cite{Hannestad:2003yd} for $2\le D\le4$, and the LHC results \cite{ATLAS:2021kxv} for $D=5,6$.
These restrictions can be converted to $M_*$ using Eq. \eqref{Planckmass}.
Denoting the scale of experimental bounds as $M_{\text{exp}}$, $M_*$ must satisfy $M_*\gtrsim M_{\text{exp}}$.\
Assuming $b_0=M^{-1}_{pl}\times10^\gamma$, where $\gamma$ is a parameter, e-folds $N_*$ is bounded by
\begin{align}
N_*\lesssim\frac{D+2}{D}\ln\left(\frac{M_{pl}}{M_{\text{exp}}}\right)-\gamma\ln10.
\end{align}
The relation between the size of extra dimensions and e-folds is summarized in Table \ref{ED_and_e-folding}.
In Table. \ref{ED_and_e-folding}, the scale of the size of extra dimension is estimated in the range of $40\le N_*\le70$.
The symbol ``$\times$'' means that $b_0$ is smaller than Planck length.
\begin{table}[htb]
\caption{The relation between the size of extra dimensions and e-folds}
\label{ED_and_e-folding}
\begin{adjustbox}{center}
\begin{tabular}{c|c|c|c|c|c|c}
Dimension $D$ & 1 & 2 & 3 & 4 & 5 & 6 \\ \hhline{=|=|=|=|=|=|=}
Bound size [$\mu$m] & 30 & $1.6\times10^{-4}$ & $2.6\times10^{-6}$ & $3.4\times10^{-7}$ & $2.1\times10^{-8}$ & $2.4\times10^{-9}$ \\
$M_{\text{exp}}$ [GeV] & $3.4\times10^8$ & $1.7\times10^6$ & $7.6\times10^4$ & $9.4\times10^3$ & $6.4\times10^3$ & $6.1\times10^3$   \\ \hline
$\gamma (60\le N_*\le70)$ & $0\le\gamma\le3$ & $\times$ &  $\times$ & $\times$ & $\times$ & $\times$  \\
$b^{-1}_0$ [GeV] & $10^{15}\sim10^{18}$ &  $\times$ &  $\times$ & $\times$ & $\times$ & $\times$ \\ \hline
$\gamma (50\le N_*\le60)$ & $4\le\gamma\le7$ & $0\le\gamma\le2$ & $\gamma\sim0$ & $\times$ & $\times$ & $\times$  \\
$b^{-1}_0$ [GeV] & $10^{11}\sim10^{14}$ & $10^{16}\sim10^{18}$ & $10^{18}$ & $\times$ & $\times$ & $\times$ \\ \hline
$\gamma (40\le N_*\le50)$ & $8\le\gamma\le9$ & $3\le\gamma\le6$ & $1\le\gamma\le5$ & $0\le\gamma\le4$ & $0\le\gamma\le3$ & $0\le\gamma\le2$ \\
$b^{-1}_0$ [GeV] & $10^{9}\sim10^{10}$ & $10^{12}\sim10^{15}$ & $10^{13}\sim10^{17}$ & $10^{14}\sim10^{18}$ & $10^{15}\sim10^{18}$ & $10^{16}\sim10^{18}$ \\
\end{tabular}
\end{adjustbox}
\end{table}

\section{Cosmological perturbation in $D+4$ dimensions} \label{CosmoPer}
We extend the cosmological perturbation in four dimensions to $D+4$ dimensions.
Cosmological perturbation in four dimensions has been reviewed in many references (for example, see \cite{Riotto:2002yw, Baumann:2009ds, Langlois:2010xc}).
Cosmological perturbation in five dimensions has been studied in \cite{Antoniadis:2023sya}.
As mentioned in \cite{Antoniadis:2023sya}, the general formalism for the cosmological perturbations in brane world theories has been studied \cite{vandeBruck:2000ju, Delfin:2023fcq, Gordon:2000hv}.

\subsection{Metric perturbations}
The perturbed metric is given by
\begin{align}
ds^2&=a^2(\tau)\Big\{-(1+2\Phi)d\tau^2+2B_i d\tau dx^i+\Big((1+2\mathcal{R})\gamma_{ij}+E_{ij}\Big)dx^i dx^j \nonumber \\
&\hspace{15mm}+2C_md\tau dy^m+2F_{im}dx^i dy^m+(b^2_0-\Xi)\eta_{mn}dy^m dy^n\Big\},
\label{D-dim_metric}
\end{align}
where $i=1,2,3$ and $m=1,2,\cdots D$ indicate the index in three and extra dimensions, respectively.
$\eta_{mn}$ is a metric for extra-dimensional space.
The perturbations $B_i, E_{ij}, C_m, F_{im}$ can be decomposed into scalar, vector, and tensor modes like
\begin{align}
B_i&=\partial_i B+B^{(V)}_i,\quad
E_{ij}=2\partial_i\partial_j E+\partial_{(i}E_{j)}+2h_{ij}, \nn \\
C_m&=\partial_m C+C^{(V)}_m,\quad
F_{im}=\partial_i F_m+F^{(V)}_{im},
\end{align}
where $\partial_{(i}E_{j)}=(\partial_i E_j+\partial_j E_i)/2$, $\partial^i h_{ij}=0$ and $h^i_{i}=0$.
Thus, there are $D+6$ scalar perturbations $(\Phi, B, \mathcal{R}, E, C, F_m, \Xi)$, $D+2$ three-dimensional vector perturbations $(B^{(V)}_i, C^{(V)}_i, F^{(V)}_{im})$, one tensor perturbation $h_{ij}$.
Since we are not interested in the vector perturbation in terms of the power spectrum, we will focus on the scalar and tensor perturbations.

Denoted $x^M$ as $x^M=(x^i,y^m)$, the spacetime coordinate is transformed by $x^M\to x^M+\xi^M$.
$\xi^M$ can be also decomposed into $\xi^M=(\xi^0,\partial^i \xi^{(S)}+\xi^{i(V)},\partial^m\xi^{(S)}+\xi^{m(V)})$.
Under these coordinate transformations, the scalar perturbations transform as
\begin{align}
&\Phi\to\Phi-\mathcal{H}\xi^0-{\xi^0}',\quad
B\to B+\xi^0-{\xi^{(S)}}',\quad
\mathcal{R}\to\mathcal{R}-\mathcal{H}\xi^0,\quad
E\to E-\xi^{(S)}, \nn \\
&C\to C+\xi^0-b^2_0{\xi^{(S)}}',\quad
F_m\to F_m-\partial_m\xi^{(S)}-b^2_0\xi_m,\quad
\Xi\to \Xi+Db^2_0\mathcal{H}'\xi^0+b^2_0\partial_m\xi^m,
\end{align}
which is derived by $\delta g_{MN}\to\delta g_{MN}-2\nabla_{(M}\xi_{N)}$.
The tensor perturbation is invariant to the coordinate transformation.

We choose the gauge $E=F_m=0$ to focus on the perturbation $\mathcal{R}$.
Moreover, we set the perturbation for the inflaton $\delta\phi=0$.\footnote{The perturbation for the inflaton transforms as $\delta\phi\to\delta\phi-\phi'\xi^0$.}
This gauge choice is called a comoving gauge.
In this gauge, $\mathcal{R}$ is interpreted as the comoving curvature perturbation.

Finally, the perturbed metric \eqref{D-dim_metric} we deal with is reduced to
\begin{align}
ds^2&=a^2(\tau)\Big\{-(1+2\Phi)d\tau^2+2(\partial_i B) d\tau dx^i+\Big((1+2\mathcal{R})\gamma_{ij}+2h_{ij}\Big)dx^i dx^j \nonumber \\
&\hspace{18mm}+2(\partial_m C)d\tau dy^m+(b^2_0-2\Xi)\eta_{mn}dy^m dy^n\Big\}.
\label{use_metric}
\end{align}

\subsection{Einstein equations}
Using Eq. \eqref{use_metric}, we write down the equations the perturbations follow from Einstein equations.
The results are obtained by performing the laborious calculations for the scalar perturbations:
\begin{itemize}
\item $00$ component
\begin{align}
&b^2_0\left((D+2)\mathcal{H}\Big[-3\mathcal{R}'+(D+3)\mathcal{H}\Phi\Big]+\Delta\Big[2\mathcal{R}+(D+2)\mathcal{H}B\Big] \right)\nn \\
&-D\Delta\Xi-\frac{D-1}{b^2_0}\Delta^{(D)}\Xi+3\Delta^{(D)}\mathcal{R}+(D+2)\mathcal{H}\Delta^{(D)}C+D(D+2)\mathcal{H}\Xi'=-b^2_0\frac{a^2}{M^2_{pl}}\delta\rho.
\label{CP:00}
\end{align}

\item $0i$ component

\begin{align}
b^2_0\Big(-4\mathcal{R}'+2(D+2)\mathcal{H}\Phi\Big)-\Delta^{(D)}(B-C)+2D\Xi'=0.
\label{CP:0i}
\end{align}

\item $0m$ component

\begin{align}
\partial_m\Delta(B-C)-6\partial_m\mathcal{R}'+2(D+2)\mathcal{H}\partial_m\Phi+2(D-1)\frac{\partial_m\Xi'}{b^2_0}=0.
\label{CP:0m}
\end{align}

\item $ii$ component

\begin{align}
&b^2_0\Big[-2\mathcal{R}''-2(D+2)\mathcal{H}\mathcal{R}'+(D+2)\mathcal{H}\Phi'+(D+2)\Big((D+1)\mathcal{H}^2+2\mathcal{H}'\Big)\Phi\Big] \nn \\
&+D\Xi''+D(D+2)\mathcal{H}\Xi'-\Delta^{(D)}\left(\mathcal{R}+\frac{\Xi}{b^2_0}\right)
=b^2_0\frac{a^2}{M^2_{pl}}\delta p.
\label{CP:ii}
\end{align}

\item $ij$ component $(i\ne j)$

\begin{align}
b^2_0\Big[B'+(D+2)\mathcal{H}B+\Phi+\mathcal{R}\Big]-D\Xi=0.
\label{CP:ij}
\end{align}

\item $im$ component

\begin{align}
b^2_0\Big[(D+2)\mathcal{H}\partial_m(B+C)+\partial_m(B'+C'+2\Phi+4\mathcal{R})\Big]-2(D-1)\partial_m\Xi=0.
\label{CP:im}
\end{align}


\item $mm$ component

\begin{align}
&b^2_0\Big[-3\mathcal{R}''-3(D+2)\mathcal{H}\mathcal{R}'+(D+2)\mathcal{H}\Phi'+(D+2)\Big((D+1)\mathcal{H}^2+2\mathcal{H}'\Big)\Phi\Big] \nn \\
&+(D-1)\Xi''+(D-1)(D+2)\mathcal{H}\Xi'+b^2_0\Delta\left(\mathcal{R}+\frac{\Xi}{b^2_0}\right)=b^2_0\frac{a^2}{M^2_{pl}}\delta p.
\label{CP:mm}
\end{align}

\item $mn$ component $(m\ne n)$

\begin{align}
C'+(D+2)\mathcal{H}C+\Phi+3\mathcal{R}-(D-2)\frac{\Xi}{b^2_0}=0.
\label{CP:mn}
\end{align}
\end{itemize}

Here, we comment on a few things.
The operators $\Delta, \Delta^{(D)}$ are defined by $\Delta=\partial_i\partial^i$, $\Delta^{(D)}=\partial_m\partial^m$, respectively.
To obtain Eqs. \eqref{CP:ii} and \eqref{CP:mm}, we use Eqs. \eqref{CP:ij} and Eq. \eqref{CP:mn}.
$\delta\rho$ and $\delta p$ are represented by
\begin{align}
\delta\rho=\delta p=-\frac{(\phi')^2}{a^2}\Phi
\end{align}
with the comoving gauge.
For $D=0,1$, Eqs. \eqref{CP:00}-\eqref{CP:mn} reproduce the equations of scalar perturbations in four dimensions and five dimensions \cite{Antoniadis:2023sya}.

We also write down the equations of tensor perturbation.
Calculating the $ij$ component of Einstein equation, which is less cumbersome than ones for the scalar perturbations, the equation of $h_{ij}$ has
\footnote{
For $D\ge2$, we can consider the extra-dimensional tensor perturbation, denoted as $e_{mn}$.
$e_{mn}$ follows the same equation of $h_{ij}$.
}
\begin{align}
h''_{ij}+(D+2)\mathcal{H}h'_{ij}-\left(\Delta+\frac{\Delta^{(D)}}{b^2_0}\right)h_{ij}=0.
\label{h:eq}
\end{align}

\subsection{Scalar perturbations}
Before investigating the power spectrum of the scalar perturbation, we mention the Fourier transformation of a $(D+4)$-dimensional field $A(\tau,x^i,x^m)$ as
\begin{align}
A(\tau,x^i, x^m)=\int d^3k\sum_{\vec{n}}A_{\vec{n}}(k,\tau)e^{i\vec{k}\cdot\vec{x}_3} e^{i\vec{n}\cdot \vec{y}_D},
\end{align}
where $k$ is a wave number, $\vec{n}=(n_,n_2\cdots n_D)$ and $n_m\in\mathbb{Z}$ for $m=1,\cdots,D$.
For a while, we consider the fields in Fourier space.
We also omit the subscript $\vec{n}$ in the fields to avoid clutter.
Thanks to the Fourier transformation, the operators $\Delta$ and $\Delta^{(D)}$ are replaced with $\Delta\to-k^2$ and $\Delta^{(D)}\to-|\vec{n}|^2$, respectively.
For convenience, we define $m_{k,n}$ as
\begin{align}
m^2_{k,n}=k^2+\frac{|\vec{n}|^2}{b^2_0}.
\end{align}

We want to know the equation that the curvature perturbation $\mathcal{R}$ follows.
However, it's hard to find this equation directly using Eqs.\eqref{CP:00}-\eqref{CP:mn}.
To find $\mathcal{R}$, we define two new variables, including $\mathcal{R}$ and $\Xi$ and derive the equations of two new variables.
Then, we find the curvature perturbation from two variables.

To define the first variable, we focus on Eqs. \eqref{CP:ii} and \eqref{CP:mm}.
These equations are reduced to
\begin{align}
\mathcal{R}''+(D+2)\mathcal{H}\mathcal{R}'+\frac{\Xi''}{b^2_0}+(D+2)\mathcal{H}\frac{\Xi'}{b^2_0}-\left(\Delta+\frac{\Delta^{(D)}}{b^2_0}\right)\left(\mathcal{R}+\frac{\Xi}{b^2_0}\right)=0.
\end{align}
The first variable is defined by
\begin{align}
\Theta\equiv\mathcal{R}+\frac{\Xi}{b^2_0},
\label{var:Theta}
\end{align}
leading to the equation of $\Theta$ as
\begin{align}
&~\Theta''+(D+2)\mathcal{H}\Theta'-\left(\Delta+\frac{\Delta^{(D)}}{b^2_0}\right)\Theta=0 \nn \\
\Rightarrow&~\Theta''+(D+2)\mathcal{H}\Theta'+m^2_{k,n}\Theta=0. \label{Theta:eq}
\end{align}

Defining the second variable is not straightforward.
Following \cite{Antoniadis:2023sya}, $\mathcal{H}\Phi$ and $B-C$ are expressed as
\begin{align}
\mathcal{H}\Phi&=\frac{1}{(D+2) m^2_{k,n}}\left[\left(2m^2_{k,n}+\frac{|\vec{n}|^2}{b^2_0}\right)\mathcal{R}'-\left(D m^2_{k,n}-\frac{|\vec{n}|^2}{b^2_0}\right)\frac{\Xi'}{b^2_0}\right], \label{HPhi} \\
B-C&=-\frac{2}{m^2_{k,n}}\left(\mathcal{R}'+\frac{\Xi'}{b^2_0}\right)=-\frac{2}{m^2_{k,n}}\Theta,
\end{align}
using Eqs. \eqref{CP:0i} and \eqref{CP:0m}.
With reference to Eq. \eqref{HPhi}, we define the second variable as
\begin{align}
\Omega\equiv\frac{1}{(D+2) m^2_{k,n}}\left[\left(2m^2_{k,n}+\frac{|\vec{n}|^2}{b^2_0}\right)\mathcal{R}-\left(D m^2_{k,n}-\frac{|\vec{n}|^2}{b^2_0}\right)\frac{\Xi}{b^2_0}\right].
\label{var:Omega}
\end{align}
By definition, $\mathcal{H}\Phi=\Omega'$ is satisfied.
Leading to the equation of $\Omega$ is redundant.
We note the following strategy: differentiate both sides of Eq. \eqref{CP:00} with respect to $\tau$ and use Eqs. \eqref{CP:ij} and \eqref{CP:mn} during its calculation.
After the calculation, we can obtain the equation of $\Omega$ as
\begin{align}
\Omega''+\left[(D+2)\mathcal{H}+\frac{2(\mathcal{H}')^2-\mathcal{H}\mathcal{H}''}{\mathcal{H}(\mathcal{H}^2-\mathcal{H}')}\right]\Omega'+m^2_{k,n}\Omega=0. \label{Omega:eq}
\end{align}
From Eqs. \eqref{var:Theta} and \eqref{var:Omega}, we obtain $\mathcal{R}$ as
\begin{align}
\mathcal{R}_{k,n}&=\Omega_{k,n}+\frac{1}{(D+2)m^2_{k,n}}\left(Dm^2_{k,n}-\frac{|\vec{n}|^2}{b^2_0}\right)\Theta_{k,n}, \label{var:R}
\end{align}
where we restore the subscript $k,n$.

\subsection{Power spectrum: scalar}\label{PS:scalar:sec3}
Performing the following variable transformations:
\begin{align}
y=a^{(D+2)/2},\quad
\theta_{k,n}=y\Theta_{k,n},\quad
z=\frac{a^{(D+2)/2}\phi'}{\mathcal{H}},\quad
\omega_{k,n}=z\Omega_{k,n},
\end{align}
on Eqs. \eqref{Theta:eq} and \eqref{Omega:eq}, we can obtain Mukhanov-Sasaki equations \cite{Mukhanov:1990me}.
The results have
\begin{align}
\theta_{k,n}''+\left(m^2_{k,n}-\frac{y''}{y}\right)\theta_{k,n}=0,\quad
\omega_{k,n}''+\left(m^2_{k,n}-\frac{z''}{z}\right)\omega_{k,n}=0.
\label{Mukhanov:eq}
\end{align}
Before looking at the solution to Eq. \eqref{Mukhanov:eq}, we compute $y''/y$ and $z''/z$ as
\begin{align}
\frac{y''}{y}
&\simeq\frac{1}{\tau^2}\left[\frac{(D+2)(D+4)}{4}+\frac{(D+2)(D+3)}{2}\epsilon\right], \\
\frac{z''}{z}&\simeq\frac{1}{\tau^2}\left[\frac{(D+2)(D+4)}{4}+\frac{(D+3)(D+6)}{2}\epsilon-(D+3)\eta\right]
\end{align}
with $\mathcal{H}\simeq-(1+\epsilon)/\tau$.
Based on these expressions, we define quantities $\nu_\theta$, $\nu_\omega$ as
\begin{align}
\frac{y''}{y}\equiv\frac{1}{\tau^2}\left(\nu^2_\theta-\frac{1}{4}\right),\quad
\frac{z''}{z}\equiv\frac{1}{\tau^2}\left(\nu^2_\omega-\frac{1}{4}\right).
\label{nu_define}
\end{align}
In detail, $\nu_\theta$ and $\nu_\omega$ are computed as
\begin{align}
\nu_\theta=\frac{D+3}{2}+\frac{D+2}{2}\epsilon,\quad
\nu_\omega=\frac{D+3}{2}+\frac{D+6}{2}\epsilon-\eta.
\end{align}
Combining Eq.\eqref{Mukhanov:eq} with Eq.\eqref{nu_define}, the equations are reduced to
\begin{align}
f''+\left[m^2_{k,n}-\left(\nu^2_f-\frac{1}{4}\right)\frac{1}{\tau^2}\right]f=0,
\label{Bessel:eq}
\end{align}
where $f=\theta_{k,n}, \omega_{k,n}$.
Eq.\eqref{Bessel:eq} is equivalent to Bessel's differential equation.
Its solution is given by
\begin{align}
f(\tau)=\sqrt{\frac{\pi}{4m_{k,n}}}\sqrt{-m_{k,n}\tau}e^{i\pi(\nu_f+1/2)/2} H^{(1)}_{\nu_f}(-m_{k,n}\tau),
\label{Bessel:sol}
\end{align}
in Bunch-Davis vacuum\footnote{In the Bunch-Davis vacuum, the solution of Eq. \eqref{Bessel:eq} approaches a plane wave solution in the short-wavelength limit.
The coefficients that appear before the Hankel function are determined thanks to these boundary conditions and normalization conditions. In detail, see \cite{Riotto:2002yw, Baumann:2009ds, Langlois:2010xc}.}.
$H^{(1)}_\nu(x)$ is the first Hankel function.
We consider the limit where the $(D+4)$-dimensional modes exit the Hubble horizon, denoted as $m_{k,n}\tau\to0$.
In this limit, the asymptotic form of the solution \eqref{Bessel:sol} can be obtained as
\begin{align}
f_k(\tau)&\rightarrow
e^{i\pi(\nu_f-\frac{1}{2})}2^{\nu_f-1}\sqrt{\frac{\tau}{\pi}}\Gamma(\nu_f)\frac{1}{(m_{k,n}\tau)^{\nu_f}}.
\label{sol:limit}
\end{align}

The power spectrum of $\mathcal{R}_{k,n}$ is defined by
\begin{align}
\mathcal{P}_{\mathcal{R}}(k)=\frac{k^3}{2\pi^2}\sum_{\vec{n}}|\mathcal{R}_{k,n}|^2,
\end{align}
which needs the summation for the Kaluza-Klein (KK) modes.
To compute the power spectrum of $\mathcal{R}_{k,n}$, we first calculate $|\mathcal{R}_{k,n}|^2$.
Using Eqs. \eqref{var:R}, \eqref{sol:limit}, and the relation $\phi'/\mathcal{H}=M_{pl}\sqrt{(D+2)\epsilon}$, $|\mathcal{R}_{k,n}|^2$ is computed as\footnote{Note that the cross term of $\theta_{k,n}$ and $\omega_{k,n}$ is absent since they are expanded by different creation and annihilation operators.}
\begin{align}
|\mathcal{R}_{k,n}|^2
&\simeq\frac{1}{a^{D+2}M^2_{pl}(D+2)\epsilon}\frac{\tau}{\pi}\left[\frac{2^{2\nu_\omega-2}\Gamma^2(\nu_\omega)}{(m_{k,n}\tau)^{2\nu_\omega}}+\frac{\epsilon}{D+2}\frac{M^2_{pl}}{m^4_{k,n}}\left(Dm^2_{k,n}-\frac{|\vec{n}|^2}{b^2_0}\right)^2\frac{2^{2\nu_\theta-2}\Gamma^2(\nu_\theta)}{(m_{k,n}\tau)^{2\nu_\theta}}\right].
\end{align}
In the slow-roll regime, $\nu_\omega\simeq\nu_\theta\simeq(D+3)/2$, the power spectrum of $\mathcal{P}_{\mathcal{R}}$ is deformed as
\begin{align}
\mathcal{P}_{\mathcal{R}}(k)&\simeq\frac{2^{D}\Gamma^2(\frac{D+3}{2})}{\pi^3(D+2)\epsilon M^2_{pl}}k^3H^{D+2}\tau^{D+3} \nn \\
&\quad\times\sum_{\vec{n}}\left[\frac{1}{(m_{k,n}\tau)^{2\nu_\omega}}+\frac{\epsilon}{D+2}\frac{M^2_{pl}}{m^4_{k,n}}\left(Dm^2_{k,n}-\frac{|\vec{n}|^2}{b^2_0}\right)^2\frac{1}{(m_{k,n}\tau)^{2\nu_\theta}}\right].
\end{align}

For convenience, we define $S_\nu(x)$ as
\begin{align}
S_\nu(x)\equiv\sum_{\vec{n}}\frac{1}{(|\vec{n}|^2+x)^\nu}. \label{Snu(x)}
\end{align}
Using $S_\nu(x)$, the following form can be represented as
\begin{align}
\sum_{\vec{n}}\frac{1}{(m_{k,n})^{2\nu}}=b^{2\nu}_0 S_\nu((b_0 k)^2).
\end{align}
Thus, $\mathcal{P}_{\mathcal{R}}$ can be expressed as
\begin{align}
\mathcal{P}_{\mathcal{R}}(k)&\simeq\frac{2^{D}\Gamma^2(\frac{D+3}{2})}{\pi^3(D+2)\epsilon M^2_{pl}}b^D_0 (b_0 k)^3H^{D+2}\tau^{D+3} 
\times\bigg[\tau^{-2\nu_\omega}S_{\nu_\omega}((b_0 k)^2) \nn \\
&\left.+\frac{\epsilon M^2_{pl}}{D+2}\tau^{-2\nu_{\theta}}\Big((D-1)^2S_{\nu_\theta}((b_0 k)^2)+2(D-1)(b_0 k)^2 S_{\nu_\theta+1}((b_0 k)^2)+(b_0 k)^4 S_{\nu_\theta+2}((b_0 k)^2)
\Big)\right].
\label{general_form_Pr}
\end{align}

Although finding the analytic form of $S_\nu(x^2)$ is hard, we can see the asymptotic form in the limit $x\ll1$ or $x\gg1$.
In the limit $x\ll1$, the contribution to the KK zero modes $\vec{n}=\vec{0}$ is the largest in $S_\nu(x^2)$.
Thus, we understand $S_\nu(x^2)$ in the $x\ll1$ limit as
\begin{align}
S_\nu(x^2)&\simeq\frac{1}{x^{2\nu}}\quad(x\ll1).\label{S:x<1}
\end{align}

On the other hand, the limit $x\gg1$ is not straightforward.
First, using the integral representation of Gamma function (Schwinger representation), $S_\nu(x)$ can be deformed as
\begin{align}
S_\nu(x^2)
&=\sum_{\vec{n}}\frac{1}{\Gamma(\nu)}\int_0^\infty dt t^{\nu-1}\exp\left[-(|\vec{n}|^2-x^2)t\right] \nn \\
&=\frac{1}{\Gamma(\nu)}\int_0^\infty dt t^{\nu-1}e^{-x^2 t}\left(\sum_{n=-\infty}^\infty e^{-n^2 t}\right)^D.
\label{S:Gamma}
\end{align}
To proceed with the calculation, we use the Poisson resummation formula:
\begin{align}
\sum_{n=-\infty}^\infty e^{-n^2 t}=\sqrt{\frac{\pi}{t}}\sum_{m=-\infty}^\infty e^{-\frac{\pi^2}{t}m^2}.
\label{Poisson}
\end{align}
Substituting Eq. \eqref{Poisson} for Eq. \eqref{S:Gamma}, one has
\begin{align}
S_\nu(x^2)
&=\frac{\pi^{D/2}}{\Gamma(\nu)}x^{D-2\nu}\sum_{\vec{n}}\int_0^\infty du u^{\frac{2\nu-D}{2}-1}\exp\left[-u-\frac{1}{4u}(2\pi x |\vec{n}|)^2\right],
\label{PRtoBessel}
\end{align}
where a change of variable $u=x^2 t$ is performed.
The integral part of Eq. \eqref{PRtoBessel} is equivalent to the integral representation of the modified Bessel function of the second kind as
\begin{align}
K_\nu(z)=\frac{1}{2}\left(\frac{z}{2}\right)^{-\nu}\int_0^\infty dt t^{\nu-1}\exp\left[-t-\frac{z^2}{4t}\right].
\end{align}
Thus, $S_\nu(x^2)$ can be expressed as
\begin{align}
S_\nu(x^2)=\frac{2\pi^{D/2}}{\Gamma(\nu)}x^{D-2\nu}\sum_{\vec{n}}\left(\frac{z}{2}\right)^{\frac{2\nu-D}{2}}K_{\frac{2\nu-D}{2}}(z),
\end{align}
where $z=2\pi x|\vec{n}|$.
Since we are interested in the behavior of Eq. \eqref{general_form_Pr}, we deal with $\nu_\omega\simeq\nu_\theta\simeq(D+3)/2$, $\nu_\theta+1\simeq(D+5)/2$, and $\nu_\theta+2\simeq(D+7)/2$ as $\nu$.
More generally, it is sufficient to find the behaviors with the $2\nu-D=3,5,7$ cases.
The modified Bessel functions of the second kind $K_{3/2}(z), K_{5/2}(z)$, and $K_{7/2}(z)$ are analytically known.
Thus, the behaviors of $(z/2)^{(2\nu-D)/2}K_{(2\nu-D)/2}(z)$ are summarized as
\begin{align}
\left(\frac{z}{2}\right)^{3/2}K_{3/2}(z)&
=\frac{\Gamma(3/2)}{2}(1+z)e^{-z}, \\
\left(\frac{z}{2}\right)^{5/2}K_{5/2}(z)&
=\frac{\Gamma(5/2)}{2}\left(1+z+\frac{z^2}{3}\right)e^{-z}, \\
\left(\frac{z}{2}\right)^{7/2}K_{7/2}(z)&
=\frac{\Gamma(7/2)}{2}\left(1+z+\frac{2z^2}{5}+\frac{z^3}{15}\right)e^{-z},
\end{align}
In the limit $x\sim z\gg1$, the terms like (polynomials of $z$)$\times e^{-z}$ can be ignored.
Therefore, the behavior of $S_\nu(x^2)$ in the limit $x\gg1$ is
\begin{align}
S_\nu(x^2)\simeq\frac{\pi^{D/2}\Gamma\left(\nu-\frac{D}{2}\right)}{\Gamma(\nu)}x^{D-2\nu}\sum_{\vec{n}} e^{-2\pi x |\vec{n}|}\simeq\frac{\pi^{D/2}\Gamma\left(\nu-\frac{D}{2}\right)}{\Gamma(\nu)}x^{D-2\nu}, \label{S:x>1}
\end{align}
where note that $e^{-2\pi x|\vec{n}|}$ with $\vec{n}\ne\vec{0}$ is exponentially suppressed.

We summarize the power spectrum of curvature perturbation.
We note that we use $a(\tau)=-1/(H\tau)$.
In the limit $b_0 k\ll1$, $\mathcal{P}_{\mathcal{R}}$ behaves as
\begin{align}
\mathcal{P}_{\mathcal{R}}(k)&\simeq\frac{2^{D}\Gamma^2\left(\frac{D+3}{2}\right)}{\pi^3(D+2)\epsilon M^2_{pl}}\frac{H^{D+2}}{k^D}\left[\left(\frac{k}{aH}\right)^{-(D+6)\epsilon+2\eta}+\frac{D}{D+2}\epsilon M^2_{pl}\left(\frac{k}{aH}\right)^{-(D+2)\epsilon}\right].
\label{PR:bk<1}
\end{align}
In the limit $b_0 k\gg1$, $\mathcal{P}_{\mathcal{R}}$ behaves as
\begin{align}
\mathcal{P}_{\mathcal{R}}(k)&\simeq\frac{2^{D-1}\pi^{\frac{D-5}{2}}\Gamma\left(\frac{D+3}{2}\right)}{(D+2)\epsilon M^2_{pl}}b^D_0 H^{D+2}\left[\left(\frac{k}{aH}\right)^{-(D+6)\epsilon+2\eta}+\frac{f(D)}{D+2}\epsilon M^2_{pl}\left(\frac{k}{aH}\right)^{-(D+2)\epsilon}\right],
\label{PR:bk>1}
\end{align}
where
\begin{align}
f(D)
&=(D-1)^2+\frac{6(D-1)}{D+3}+\frac{15}{(D+3)(D+5)}.
\end{align}
We reproduce the power spectrum in four dimensions in the result \eqref{PR:bk>1} with $D=0$.
Similarly, we replicate the power spectrum in five dimensions \cite{Antoniadis:2023sya} from Eqs. \eqref{PR:bk<1} and \eqref{PR:bk>1} with $D=1$.

\subsection{Power spectrum: tensor}
The equation of $h_{ij}$ is the same as Eq. \eqref{Theta:eq} in Fourier space, denoted as $h_{k,n:ij}$.
Performing the variable transformation $v_{k,n:ij}=yh_{k,n:ij}$, where $y=a^{(D+2)/2}$, we obtain
\begin{align}
v''_{k,n:ij}+\left(m^2_{k,n}-\frac{y''}{y}\right)v_{k,n:ij}=0.
\end{align}
The solution of this equation is given by Eq. \eqref{Bessel:sol} with $f=v_{k,n:ij}$.

The power spectrum of $\mathcal{P}_{h}$ is defined by
\begin{align}
\mathcal{P}_{h}=\frac{k^3}{2\pi^2}\frac{2\cdot4}{M^2_{pl}}\sum_{\vec{n}}|h_{ij}|^2,
\end{align}
where the factor of 2 comes from the two polarizations and $4/M^2_{pl}$ is a normalization factor.
The results have
\begin{align}
\mathcal{P}_h(k)&\simeq\begin{cases}
\dfrac{2^{D+3}\Gamma^2\left(\frac{D+3}{2}\right)}{\pi^3 M^2_{pl}}\dfrac{H^{D+2}}{k^D}\left(\dfrac{k}{aH}\right)^{-(D+2)\epsilon}\quad(b_0 k\ll1) \\\\
\dfrac{2^{D+2}\pi^{\frac{D-5}{2}}\Gamma\left(\frac{D+3}{2}\right)}{M^2_{pl}}b^D_0H^{D+2}\left(\dfrac{k}{aH}\right)^{-(D+2)\epsilon}\quad(b_0 k\gg1).
\end{cases}
\label{tensorPS}
\end{align}
The results \eqref{tensorPS} also reproduce the power spectrum of tensor perturbation in four dimensions and five dimensions.

\subsection{Spectral index and tensor-scalar ratio}
We can lead to the spectral index $n_s$, tensor-scalar ratio $r$, and the scalar amplitude $A_s$ from the power spectrum of scalar and tensor perturbations in $D+4$ dimensions.
Their definitions are given by
\begin{align}
\mathcal{P}_{\mathcal{R}}=A_s\left(\frac{k}{k_*}\right)^{n_s-1},\quad
r=\frac{\mathcal{P}_{h}}{\mathcal{P}_{\mathcal{R}}}.
\end{align}
According to Planck 2018 results \cite{Planck:2018vyg, Planck:2018jri}, the spectral index, tensor-scalar ratio, and the scalar amplitude are reported as
\begin{align}
n_s=0.9649\pm0.0042,\quad
r<0.10,\quad
\ln(10^{10} A_s)=3.044\pm0.014,
\end{align}
where the bound of $r$ is given at the pivot scale $k_*=0.002$ Mpc$^{-1}$.

Note that $\epsilon\simeq\epsilon_V\ll1$, the spectral index can be read from Eqs. \eqref{PR:bk<1} and \eqref{PR:bk>1} as
\begin{align}
n_s&=1-(D+6)\epsilon_V+2\eta_V-D,\quad(b_0 k\ll1) \\
n_s&=1-(D+6)\epsilon_V+2\eta_V \hspace{15mm}(b_0 k\gg1). \label{ns}
\end{align}
The spectral index is not close to one in the limit $b_0 k\ll1$ with $D\ge1$. 
This limit means that there is no scale invariance. 
The spectral index in the limit $b_0 k\gg1$ has a possibility to be compatible with Planck 2018 result.
Hereafter, we assume the $b_0 k\gg1$ case.

From Eqs. \eqref{PR:bk>1} and \eqref{tensorPS}, tensor-scalar ratio can be computed as
\begin{align}
r=8(D+2)\epsilon_V.\label{ratio}
\end{align}
By definition, the scalar amplitude can be obtained as
\begin{align}
A_s=\dfrac{2^{D-1}\pi^{\frac{D-5}{2}}\Gamma\left(\frac{D+3}{2}\right)}{(D+2)\epsilon M^2_{pl}}b^D_0 H^{D+2}.
\label{amp:Ddim}
\end{align}
Using Eqs.\eqref{FD:ACC} and \eqref{EandP} in the slow-roll regime, the inflaton potential $V(\phi)$ can be represented as
\begin{align}
V&=\frac{(D+2)(D+3)}{8}\left(\frac{r A_s M^{D+4}_{pl}}{\pi^{\frac{D-5}{2}}\Gamma\left(\frac{D+3}{2}\right)b^D_0}\right)^{\frac{2}{D+2}}. \label{inflaton:pt}
\end{align}
Combining $r<0.10$ with Eq.\eqref{inflaton:pt}, the upper bound of the energy scale of inflaton can be derived.
Taking $D=1$ and $N_*=60$ as an example of the dark dimension ($b^{-1}_0\sim10^{15}$\,GeV), the energy scale of inflaton is the order of $10^{16}$\,GeV.

Here, we comment on the orbifold.
If we consider the orbifold as the extra-dimensional space like $S^1/\mathbb{Z}_2\times S^1/\mathbb{Z}_2\times S^1/\mathbb{Z}_2$, the asymptotic behaviors of Eqs. \eqref{S:x<1} and \eqref{S:x>1} are slightly changed.
In appendix \ref{orbifold}, we discuss the behaviors of $S_\nu(x^2)$ with orbifold.
As a result, the spectral index and tensor-scalar ratio are independent of the effects of the orbifold.
However, the scalar amplitude is affected by the orbifold by a factor of $1/2^D$.
The factor $1/2^D$ is interpreted with the number of the KK spectrum projected out.

\section{Examples}\label{ex}
In the previous section, we obtained the spectral index $n_s$ and tensor-scalar ratio $r$ in $D+4$ dimensions.
In this section, we investigate five inflationary models, motivated by Fig. 8 from \cite{Planck:2018jri}.
If you want to see other models, see \cite{Martin:2013tda}.

\subsection{Chaotic inflation}
Chaotic inflation \cite{Linde:1983gd} is the simplest model, which has the power law potential as
\begin{align}
V(\phi)=\frac{\lambda}{n}\phi^n,
\end{align}
where $n>0$ and $\lambda$ is a coupling.
Calculating the slow-roll parameters, we have
\begin{align}
\epsilon_V=\frac{n^2(D+2)}{4}\left(\frac{M_{pl}}{\phi}\right)^2,\quad
\eta_V=\frac{n(n-1)(D+2)}{2}\left(\frac{M_{pl}}{\phi}\right)^2.
\label{slow:chaotic}
\end{align}
Performing the integral in the number of e-folds, $N_*$ becomes
\begin{align}
N_*=\frac{1}{n(D+2)M^2_{pl}}(\phi^2_*-\phi^2_{\text{end}}).
\end{align}
Taking into account the condition when inflation ends, $\phi_*$ has the form
\begin{align}
\phi_*
=\frac{M_{pl}}{2}\sqrt{n(D+2)(4N_*+n)}. \label{phi*:chaotic}
\end{align}
From Eqs. \eqref{slow:chaotic} and \eqref{phi*:chaotic}, $n_s$ and $r$ are represented as
\begin{align}
n_s=1-\frac{n(D+2)+4}{4N_*+n},\quad
r=8(D+2)\frac{n}{4N_*+n}.
\label{ns-r:chaos}
\end{align}

\begin{figure}[ht]
\begin{center}
\includegraphics[width=8cm]{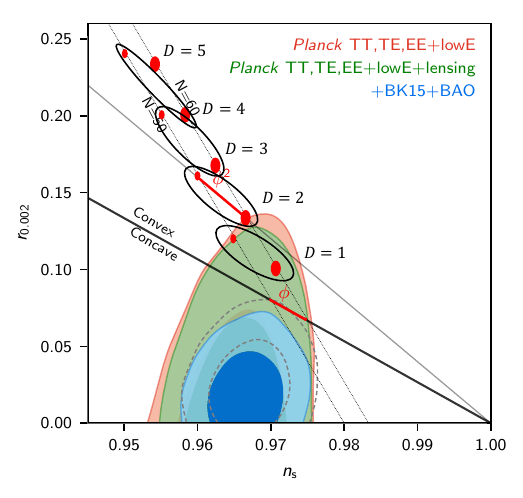}
\end{center}
\vspace*{-5mm}
\caption{Chaotic inflation with $n=1$.
Small and large red circles represent $N_*=50$ and $N_*=60$, respectively.
$n_s$-$r$ plot is taken from \cite{Planck:2018vyg}.
}
\label{CI}
\end{figure}

Taking $n=2$ and $D=1$ as an example, we get $n_s=0.9505$ and $r=0.24$ at $N_*=50$, and $n_s=0.9587$ and $r=0.20$ at $N_*=60$.
These results are excluded from Planck 2018 results.
More generally, the chaotic inflationary model with $n\ge2$ and $D\ge1$ is excluded.
The remaining possibility is the case of $n\le1$ and $D\ge1$.
Fig. \ref{CI} shows the results of chaotic inflation with $n=1$ and $D\ge1$.
As can be seen from Fig. \ref{CI}, only the case with $D=1$ is within the allowed region at CL 95\%,
which is computed by $n_s=0.9652$ and $r=0.12$ at $N_*=50$, and $n_s=0.9710$ and $r=0.10$ at $N_*=60$.

Dotted lines in the original $n_s$-$r$ plot from \cite{Planck:2018vyg} show the trajectories of the approximately constant $N_*$ in chaotic inflation.
These dotted lines can be represented as
\begin{align}
r\simeq -8n_s+8\left(1-\frac{1}{N_*}\right).
\label{dotted:line}
\end{align}
Our results seem to follow these dotted lines.
This is not a coincidence.
Eq. \eqref{slow:chaotic} with $n=1$ can be approximately reduced to Eq. \eqref{dotted:line}.

\subsection{Natural inflation}
The idea of natural inflation \cite{Freese:1990rb, Savage:2006tr} is that the inflaton is regarded as pseudo Nambu Goldstone boson.
In this model, the potential has
\begin{align}
V(\phi)=V_0\left(1+\cos\left(\frac{\phi}{f}\right)\right),
\end{align}
where $f$ is a spontaneous breaking scale.
Calculating the slow-roll parameters, one has
\begin{align}
\epsilon
=\frac{D+2}{4}\left(\frac{M_{pl}}{f}\right)^2\frac{1-\cos\frac{\phi}{f}}{1+\cos\frac{\phi}{f}},\quad
\eta=-\frac{D+2}{2}\left(\frac{M_{pl}}{f}\right)^2\frac{\cos\frac{\phi}{f}}{1+\cos\frac{\phi}{f}}.
\end{align}
The number of e-folds can be obtained as
\begin{align}
N_*
=\frac{2f^2}{(D+2)M^2_{pl}}\ln\frac{1-\cos\frac{\phi_{\text{end}}}{f}}{1-\cos\frac{\phi_*}{f}}.
\end{align}
 From $N_*$ and the condition $\epsilon(\phi_{\text{end}})=1$, $\cos(\phi_{*}/f)$ is derived as
\begin{align}
\cos\frac{\phi_*}{f}=1-\frac{2}{\frac{D+2}{4}\left(\frac{M_{pl}}{f}\right)^2+1}\exp\left[-\frac{D+2}{2}\left(\frac{M_{pl}}{f}\right)^2N_*\right]
\label{NI:phi_initial}
\end{align}
Thus, we obtain the spectral index and tensor-scalar ratio as
\begin{align}
n_s&=1-\frac{D+2}{4}\left(\frac{M_{pl}}{f}\right)^2\frac{(D+6)-(D+2)\cos\frac{\phi_*}{f}}{1+\cos\frac{\phi_*}{f}},\\
r&=2(D+2)^2\left(\frac{M_{pl}}{f}\right)^2\frac{1-\cos\frac{\phi_*}{f}}{1+\cos\frac{\phi_*}{f}}
\end{align}

 	\begin{figure}[htbp]
	  \begin{minipage}[b]{0.5\linewidth}
	    \centering
	    \includegraphics[keepaspectratio, scale=0.5]{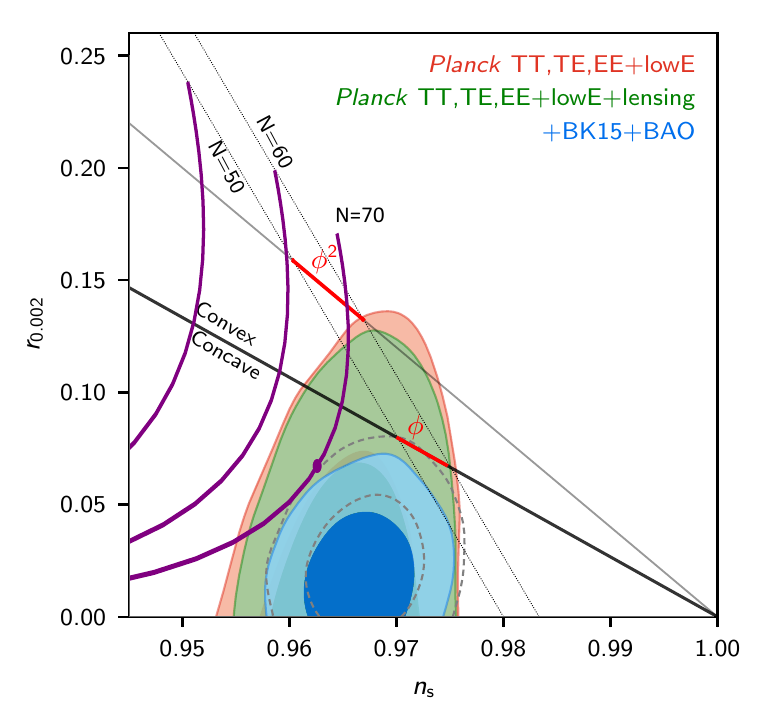}
	    \subcaption{Natural inflation: $D=1$}
	  \end{minipage}
	  \begin{minipage}[b]{0.5\linewidth}
	    \centering
	    \includegraphics[keepaspectratio, scale=0.5]{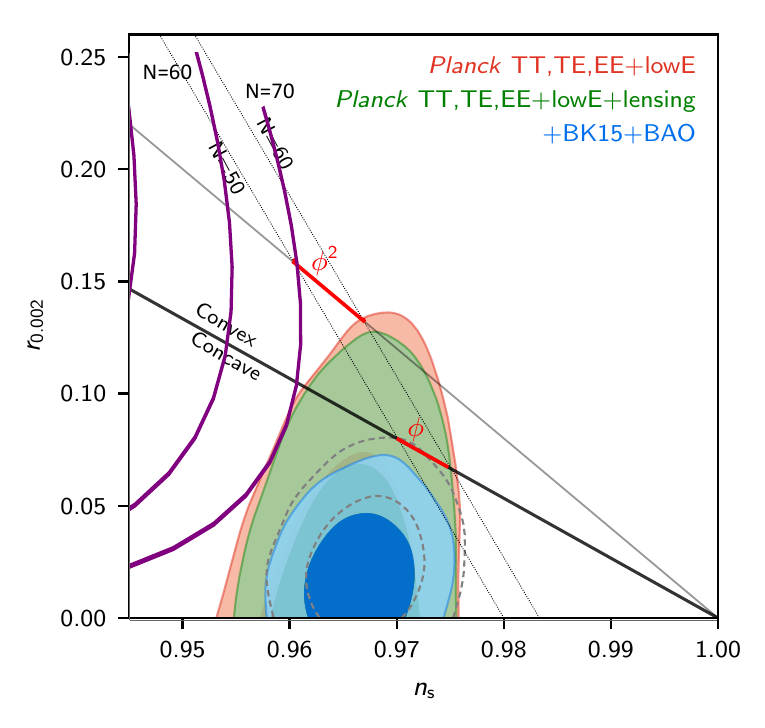}
	    \subcaption{Natural inflation: $D=2$}
	  \end{minipage}
	  \caption{Natural inflation with $D=1,2$.
	  Our results represent purple lines.
	  $n_s$-$r$ plot is taken from \cite{Planck:2018vyg}.}
	  \label{natural_inflation_results}
	  \end{figure}

We show the results of natural inflation with $D=1,2$ in Fig. \ref{natural_inflation_results}.
As $M_{pl}/f$ gets larger, the purple curves go down to the left since both $n_s$ and $r$ decrease.
In both cases with $D=1,2$, natural inflation is excluded.
More generally, the natural inflation model with $D\ge1$ and $N_*\le60$ is excluded.
Only the case with $D=1$ and $60<N_*\le70$ is within the allowed region at CL 95\%.
Taking $f=8M_{pl}$ as a representative value (the dot in Fig. \ref{natural_inflation_results} (a)), we obtain $n_s=0.9627$ and $r=0.067$ at $N_*=70$.

\subsection{Quartic hilltop inflation}
Hilltop inflation \cite{Linde:1981mu, Albrecht:1982wi, Boubekeur:2005zm} has the following potential:
\begin{align}
V(\phi)=V_0\left(1-\frac{\phi^p}{\mu^p}+\cdots\right),
\end{align}
where $p\ge3$ and $\mu$ is a parameter with a mass dimension, and the ellipsis indicates higher-order terms.
In this analysis, we do not take the higher-order terms into account.
Slow-roll parameters are calculated as
\begin{align}
\epsilon=\frac{D+2}{4}M^2_{pl}\frac{p^2}{\phi^2}\left(\frac{\frac{\phi^p}{\mu^p}}{1-\frac{\phi^p}{\mu^p}}\right)^2,\quad
\eta=-\frac{D+2}{2}M^2_{pl}\frac{p(p-1)}{\phi^2}\frac{\frac{\phi^p}{\mu^p}}{1-\frac{\phi^p}{\mu^p}}.
\end{align}
The specific form of $N_*$ has
\begin{align}
N_*=\left.\frac{\phi^2}{p(D+2)M^2_{pl}}\left(1+\frac{2}{p-2}\left(\frac{\phi}{\mu}\right)^{-p}\right)\right|_{\phi_{\text{end}}}^{\phi_*}.
\end{align}
With reference to \cite{Planck:2018jri}, we investigate the quartic hilltop inflation, which is the case with $p=4$.
Here, $\mu^4$ is replace with $M^4_{pl}/\lambda$.
Note that the coupling $\lambda$ is dimensionless.
We must pay attention to the contribution to $\phi_{\text{end}}$ \cite{Dimopoulos:2020kol}.
Taking this contribution into account, the slow-roll parameters are represented as
\begin{align}
\epsilon=\frac{D+2}{4N_i}\frac{z^2f^3(z)}{[1-zf(z)]^2},\quad
\eta=-\frac{3(D+2)}{4N_i}\frac{zf(z)}{1-zf(z)},
\label{slow-roll:hilltop}
\end{align}
where $z$ and $f(z)$ are defined as
\begin{align}
z=16\lambda N^2,\quad
f(z)=1-\sqrt{1-\frac{1}{z}},
\end{align}
and $N$ is defined as
\begin{align}
N\equiv\frac{1}{4(D+2)\lambda}\left[\left(\frac{\phi_*}{M_{pl}}\right)^{-2}+\lambda\left(\frac{\phi_*}{M_{pl}}\right)^{2}\right].
\end{align}
The spectral index and tensor-scalar ratio are calculated as
\begin{align}
n_s&=1-\frac{D+2}{4N_i}\frac{zf(z)}{[1-zf(z)]^2}\Big(2(D+3)zf(z)-D\Big), \label{ns:hilltop} \\
r&=512(D+2)^2\lambda^2\frac{N^3_i f^3(z)}{[1-zf(z)]^2}, \label{r:hilltop}
\end{align}
using Eq. \eqref{slow-roll:hilltop}.
Eqs. \eqref{ns:hilltop} and \eqref{r:hilltop} with $D=0$ and $\lambda\ll1$ reproduce chaotic inflation with $D=0$ and $n=1$, that is $n_s=1-3/(2N_*)$ and $r=4/N_*$.

	\begin{figure}[htbp]
	  \begin{minipage}[b]{0.5\linewidth}
	    \centering
	    \includegraphics[keepaspectratio, scale=0.5]{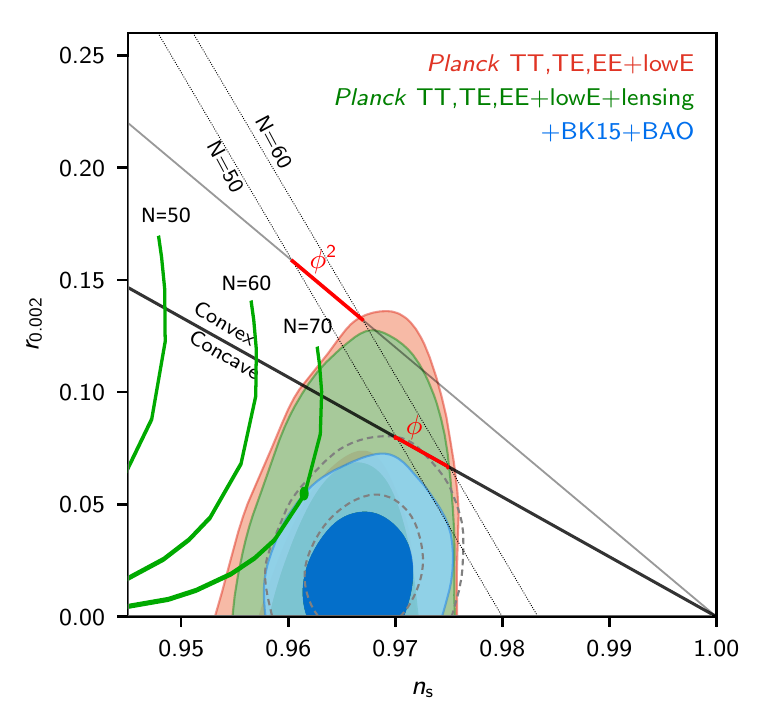}
	    \subcaption{Quartic hilltop inflation: $D=1$}
	  \end{minipage}
	  \begin{minipage}[b]{0.5\linewidth}
	    \centering
	    \includegraphics[keepaspectratio, scale=0.5]{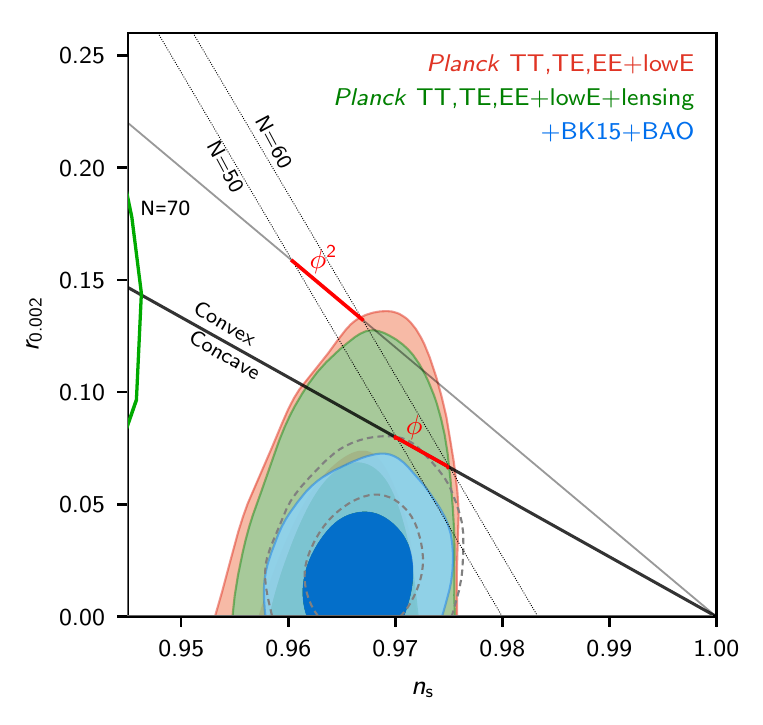}
	    \subcaption{Quartic hilltop inflation: $D=2$}
	  \end{minipage}
	  \caption{Quartic hilltop inflation with $D=1,2$.
	  Our results represent green lines.
	  $n_s$-$r$ plot is taken from \cite{Planck:2018vyg}.}
	  \label{hilltop_inflation_results}
	  \end{figure}

We show the results of quartic hilltop inflation with $D=1,2$ in Fig. \ref{hilltop_inflation_results}.
As $\lambda$ increases to $\lambda\sim10^{-2}$, the green curves go down to the left.
As in the natural inflation model, the quartic hilltop inflation with $D=1,2$ is excluded.
More generally, quartic hilltop inflation with $D\ge1$ and $N_*\le60$ is excluded.
Only the case with $D=1$ and $60<N_*\le70$ is within the allowed region at CL 95\%.
Taking $\lambda=10^{-6}$ as a representative value (the dot in Fig. \ref{hilltop_inflation_results} (a)), we obtain $n_s=0.9616$ and $r=0.054$ at $N_*=70$.

\subsection{Spontaneously broken SUSY model}
Inflation with spontaneously broken SUSY \cite{Dvali:1994ms} is motivated by the low-scale SUSY model.
Its potential is given by
\begin{align}
V(\phi)=V_0\left[1+\alpha_h\ln\frac{\phi}{M_{pl}}\right],
\end{align}
where $\alpha_h$ is a parameter.
This parameter is roughly the loop factor.
Calculating the slow-roll parameters, we have
\begin{align}
\epsilon=\frac{D+2}{4}\left(\frac{\alpha_hM_{pl}}{\phi}\right)^2\frac{1}{(1+\alpha_h\ln\frac{\phi}{M_{pl}})^2},\quad
\eta=-\frac{D+2}{2}\left(\frac{M_{pl}}{\phi}\right)^2\frac{\alpha_h}{1+\alpha_h\ln\frac{\phi}{M_{pl}}}.
\end{align}
Performing the integral in $N_*$, we obtain
\begin{align}
N_*=\left.\frac{2}{(D+2)\alpha_h M^2_{pl}}\left\{\frac{1}{2}\phi^2\left(1+\alpha_h\ln\frac{\phi}{M_{pl}}\right)-\frac{\alpha_h}{4}\phi^2\right\}\right|_{\phi_{\text{end}}}^{\phi_*}.
\end{align}
$\phi_*$ is approximately represented as
\begin{align}
\phi_*\simeq\sqrt{\frac{(D+2)\alpha_h N_*}{1-\frac{1}{2}\alpha_h}}M_{pl}.
\end{align}
Therefore, $n_s$ and $r$ are
\begin{align}
n_s&=1-\frac{1}{N_*}\frac{\left(1-\frac{\alpha_h}{2}\right)}{(1+\alpha_h\ln\frac{\phi_*}{M_{pl}})^2}\left(\frac{D+6}{4}\alpha_h+1+\alpha_h\ln\frac{\phi_*}{M_{pl}}\right), \\
r&=2(D+2)\alpha_h\frac{\left(1-\frac{\alpha_h}{2}\right)}{(1+\alpha_h\ln\frac{\phi_*}{M_{pl}})^2}.
\end{align}

In this model, the tensor-scalar ratio $r$ predicts a very small value as $\alpha_h$ decreases.
However, regardless of the influence of dimensions, $n_s\approx0.980$ at $N_*=50$ and $n_s=0.983$ at $N_*=60$ are obtained if we take $\alpha_h\le10^{-1}$.
Therefore, the inflationary model with spontaneously broken SUSY is not revived.

\subsection{$R^2$ inflation}
$R^2$ inflation \cite{Starobinsky:1980te} is the most supported model in the inflationary models by Planck 2018 results.
Starting from $(D+4)$-dimensional theory, we derive the potential as
\begin{align}
V(\phi)
&=V_0 \exp\left[\frac{D}{\sqrt{(D+2)(D+3)}}\frac{\phi}{M_{pl}}\right]\left(1-\exp\left[-\sqrt{\frac{D+2}{D+3}}\frac{\phi}{M_{pl}}\right]\right)^2.
\label{SI:pot}
\end{align}
Note that this potential with $D=0$ reproduces the original Starobinsky type potential $V\propto(1-e^{-\sqrt{2/3}\phi/M_{pl}})^2$.
The derivation of Eq. \eqref{SI:pot} is summarized in appendix \ref{derivation}.
The slow-roll parameters in this model are calculated as
\begin{align}
\epsilon&=\frac{\left((D+4)+D\exp\left[\sqrt{\frac{D+2}{D+3}}\frac{\phi}{M_{pl}}\right]\right)^2}{4(D+3)}\left(1-\exp\left[\sqrt{\frac{D+2}{D+3}}\frac{\phi}{M_{pl}}\right]\right)^{-2}, \\
\eta&=\frac{1}{2(D+3)}\left((D+4)^2-8\exp\left[\sqrt{\frac{D+2}{D+3}}\frac{\phi}{M_{pl}}\right]+D^2\exp\left[2\sqrt{\frac{D+2}{D+3}}\frac{\phi}{M_{pl}}\right]\right) \nn \\
&\qquad\times\left(1-\exp\left[\sqrt{\frac{D+2}{D+3}}\frac{\phi}{M_{pl}}\right]\right)^{-2}.
\end{align}
From the condition when inflation ends, $\phi_{\text{end}}/M_{pl}$ is derived as
\begin{align}
\frac{\phi_{\text{end}}}{M_{pl}}=\sqrt{\frac{D+3}{D+2}}\ln\left(\frac{D+6+4\sqrt{D+3}}{6-D}\right),
\label{SID:xend}
\end{align}
which is solved under the condition of $-2<D<6$.

Considering $D>0$, the slow-roll parameters converge to finite values in the limit $\phi/M_{pl}\to\infty$.
The results have
\begin{align}
\epsilon\to\frac{D^2}{4(D+3)},\quad
\eta\to\frac{D^2}{2(D+3)}\quad
\left(\frac{\phi}{M_{pl}}\to\infty\right).
\end{align}
These finite values are the lowest in $\phi/M_{pl}>0$.
Thus, the spectral index and tensor-scalar ratio are summarized in
\begin{align}
n_s\to1-\frac{D^2(D+2)}{4(D+3)},\quad
r\to\frac{2D^2(D+2)}{D+3},
\end{align}
in the limit $\phi/M_{pl}\to\infty$.
Even in the $D=1$ case, we get $n_s=0.8125$ and $r=1.5$.
Thus, $R^2$ inflation with expanded extra dimensions is excluded.

\section{Conclusion}\label{discuss}
We have considered the $(D+4)$ dimensional inflation, which causes three-dimensional non-compact space and $D$-dimensional extra space to expand uniformly.
In our setup, we have defined the two potential slow-roll parameters by Eq. \eqref{SRparameters} and e-folds by Eq. \eqref{e-folds}.
We have also computed the cosmological perturbations in $D+4$ dimensions.
From the power spectrums of scalar perturbation and tensor perturbation, we have read the spectral index and tensor-scalar ratio as $n_s=1-(D+6)\epsilon_V+2\eta_V$ and $r=8(D+2)\epsilon_V$ in the limit $b_0 k\gg1$.
Finally, we analyze five typical inflationary models and compare the results with Planck 2018 constraint.

Of the five inflationary models, chaotic inflation ($ n\ge2$), natural inflation, and quartic hilltop inflation with $ D\ge1$ and $ N_*\le60$ are excluded.
As seen from Fig. \ref{CI}, the chaotic inflation with $n\le1$ and $D=1$ is within the allowed region at CL 95\%.
In Fig. \ref{natural_inflation_results}, the natural inflation with $D=1$ and $60<N_*\le70$ is within the allowed region at CL 95\%.
In Fig. \ref{hilltop_inflation_results}, the quartic hilltop inflation with $D=1$ and $60<N_*\le70$ is within the allowed region at CL 95\%.
We have also analyzed the remaining two cases: inflation with spontaneously broken SUSY and $R^2$ inflation.
Even if we consider the contributions from the extra dimensions, inflation with spontaneously broken SUSY remains excluded.
On the other hand, analyzing the $R^2$ inflation in $D+4$ dimensions, we have found that $R^2$ inflation is excluded in our setup.

In our analysis, there are two perspectives.
First, Planck 2018 constraint supports that the universe with $D$-dimensional compact space only expands in three-dimensional non-compact space, not $D$-dimensional compact space.
Although the superstring theory, a candidate for quantum gravity, predicts that our universe has ten dimensions, it is a mystery why the expansion of extra-dimensional space is not favored.
Second, our results suggest that only one extra dimension can be expanded, as mentioned in \cite{Anchordoqui:2023etp}.
The expansion of one extra dimension would imply the dark dimension proposal \cite{Montero:2022prj}.
It would be very interesting to explore these perspectives further.
We will come back to these issues in the future.

Although inflation with spontaneously broken SUSY and $R^2$ inflation are excluded from our results, it would be worthwhile to reconsider these two models theoretically and phenomenologically.
For the inflation with spontaneously broken SUSY, we applied the four-dimensional potential of the SUSY model for simplicity.
With the appropriate potential in higher dimensions, this model may be revived.
For $R^2$ inflation, the model may satisfy Planck 2018 constraint if we add $R^n$ with $n\ge3$ to the $(D+4)$-dimensional action \cite{Otero:2017thw}.
Revisiting two inflationary models in higher-dimensional uniform inflation is left for future work.

It would also be interesting to revisit another inflationary model in the higher-dimensional uniform inflation.
One example is extranatural inflation \cite{Arkani-Hamed:2003xts, Arkani-Hamed:2003wrq}.
\footnote{Another inflationary models from higher-dimensional gauge theories are studied in \cite{Inami:2010ke, Abe:2014eia, Furuuchi:2014cwa, Abe:2015bba, Hirose:2021xbs}.}
In this model, the inflaton is identified with the extra component of gauge fields.
This model may be able to simultaneously consider the higher-dimensional uniform inflation and the radion stabilization we do not discuss.
Moduli inflation \cite{Conlon:2005jm, Ben-Dayan:2008fhy, Kobayashi:2016mzg, Abe:2023ylh, Ding:2024neh} is also interesting inflationary model.
In this model, the modulus field is a candidate of the inflaton.
Modular symmetry and the modulus are well-motivated by higher-dimensional theories such as the superstring theory.
Exploring the extranatural inflation and the moduli inflation is also left for future work.

\section*{Acknowledgments}
The author would like to thank Yuichi Koga and Takahumi Kai for the valuable discussions.

\appendix
\section{Power spectrum with orbifold}\label{orbifold}
In this appendix, we consider the orbifold $S^1/\mathbb{Z}_2\times S^1/\mathbb{Z}_2\times\cdots S^1/\mathbb{Z}_2$ for simplicity.
The behaviors of $S_\nu(x)$ defined by Eq. \eqref{Snu(x)} would be changed.
If we consider that all KK zero modes are projected out, the summation in $S_\nu(x)$ is changed as
\begin{align}
S_\nu(x)\equiv\sum_{\vec{n}\ne\vec{0}}\frac{1}{(|\vec{n}|^2+x)^\nu}.
\end{align}
Let's consider the $S_\nu(x^2)$ in the limit $x\ll1$ and $x\gg1$.
First, we consider the limit $x\ll1$.
In this limit, $S_\nu(x^2)$ becomes a constant.
In fact, $S_\nu(x^2)$ with $D=1$ and $x\ll1$ can be understood by zeta function $\zeta(2\nu)$.
Thus, the behavior of $S_\nu(x)$ is understood as
\begin{align}
S_\nu(x^2)\simeq c_D\quad(x\ll1),
\end{align}
where $c_D$ is a constant.

Next, we consider the limit $x\gg1$.
The basic method is the same as in section \ref{PS:scalar:sec3}.
The only difference from section \ref{PS:scalar:sec3} is that we use the following Poisson resummation formula:
\begin{align}
\sum_{n=1}^\infty e^{-n^2 t}=-\frac{1}{2}+\frac{1}{2}\sqrt{\frac{\pi}{t}}\sum_{m=-\infty}^\infty e^{-\frac{\pi^2}{t}m^2}.
\label{Poisson:orbifold}
\end{align}
Using Eqs. \eqref{S:Gamma} and \eqref{Poisson:orbifold}, we obtain
\begin{align}
S_\nu(x^2)&=\frac{1}{\Gamma(\nu)}\int_0^\infty dt t^{\nu-1}e^{-x^2 t}\left(-\frac{1}{2}+\frac{1}{2}\sqrt{\frac{\pi}{t}}\sum_{n=-\infty}^\infty e^{-\frac{\pi^2}{t}n^2}\right)^D \nn \\
&=\frac{1}{\Gamma(\nu)}\int_0^\infty dt t^{\nu-1}e^{-x^2 t}\left[\left(-\frac{1}{2}\right)^D+\cdots+\left(\frac{1}{2}\sqrt{\frac{\pi}{t}}\sum_{n=-\infty}^\infty e^{-\frac{\pi^2}{t}n^2}\right)^D\right]. \label{expand:S} 
\end{align}
Of all terms in Eq. \eqref{expand:S},  we focus on the first term and calculate
\begin{align}
S_\nu(x^2)&=\frac{(-1)^D}{2^D\Gamma(\nu)}\int_0^\infty dt t^{\nu-1}e^{-x^2 t}=\left(-\frac{1}{2}\right)^D\frac{1}{x^{2\nu}}
\end{align}
The contribution to $1/x^{2\nu}$ is smaller than the contribution to $1/x^{2\nu-D}$ in the limit $x\gg1$.
The same discussion applies to the other terms except for the last term in Eq. \eqref{expand:S}.
Therefore, we conclude the behaviors of $S_\nu(x^2)$ with orbifold as
\begin{align}
S_\nu(x^2)\simeq\frac{\pi^{D/2}\Gamma\left(\nu-\frac{D}{2}\right)}{2^D\Gamma(\nu)}x^{D-2\nu}\quad(x\gg1)
\label{S:x>1:orbifold}
\end{align}
The difference between Eq. \eqref{S:x>1} and Eq. \eqref{S:x>1:orbifold} is the factor of $1/2^D$.
This factor can be interpreted with the number of the KK spectrum projected out.

\section{The derivation of the $R^2$ inflationary potential in $D+4$ dimensions}\label{derivation}
Following \cite{Otero:2017thw}, we derive the $R^2$ inflationary potential in $D+4$ dimensions.
The action is given by
\begin{align}
S=\frac{M^{D+2}_{*}}{2}\int d^4x d^{D}y \sqrt{-g}\left(R+\frac{R^2}{M^2}\right)\equiv\frac{M^{D+2}_{*}}{2}\int d^{4}x d^Dy \sqrt{-g} f(R),
\end{align}
where $M$ is a constant with the dimension of a mass.
Here, we define $F(R)$ by
\begin{align}
F(R)\equiv\frac{\partial f}{\partial R}=1+\frac{2R}{M^2}.
\end{align}
Using $F(R)$, the action can be deformed as
\begin{align}
S=\frac{M^{D+2}_{*}}{2}\int d^{4}x d^Dy\sqrt{-g}\left[FR-\left(\frac{R}{M}\right)^2\right].
\end{align}

Performing a conformal transformation as
\begin{align}
\tilde{g}_{MN}=\Omega^2(x)g_{MN}=e^{2\omega(x)}g_{MN},
\end{align}
we obtain \cite{Otero:2017thw}
\begin{align}
\sqrt{-g}&=\Omega^{-(D+4)}\sqrt{-\tilde{g}}, \\
R&\supset\Omega^2\Big[\tilde{R}-(D+3)(D+2)\tilde{g}^{MN}\partial_M\omega\partial_N\omega\Big], \label{R:conformal}
\end{align}
where the linear term of $\Omega$ or $\omega$ is ignored in Eq. \eqref{R:conformal}.
The action performed by conformal transformation is represented as
\begin{align}
S&=\int d^{4}x d^Dy \sqrt{-\tilde{g}}\left[
\frac{M^{D+2}_{*}}{2}\frac{F}{\Omega^{D+2}}\Big(\tilde{R}-(D+3)(D+2)\tilde{g}^{MN}\partial_M\omega\partial_N\omega\Big)
-U(\omega)
\right], \label{action:conformaled} \\
U(\omega)&=\frac{M^{D+2}_{*}}{2\Omega^{D+4}}\left(\frac{R}{M}\right)^2.
\end{align}
Denoting the metric in four dimensions as $\tilde{g}^{4D}$, $\sqrt{-g}$ can be factorized into $\sqrt{-g}=\sqrt{-g^{4D}}b^D$.
Using Eq. \eqref{e-folds_M*}, the action can be rewritten as
\begin{align}
S&=\int d^{4}x d^Dy \sqrt{-\tilde{g}^{4D}}\left[
\frac{M^{2}_{pl}}{2}\frac{F}{\Omega^{D+2}}\Big(\tilde{R}-(D+3)(D+2)\tilde{g}^{MN}\partial_M\omega\partial_N\omega\Big)
-V(\omega)
\right], \label{action:conformaled2} \\
V(\omega)&=\frac{1}{2\Omega^{D+4}}\left(\frac{M_{pl}}{M}\right)^2R^2.
\end{align}

To make the kinetic terms of $\tilde{R}$ and $\omega$, we require
\begin{align}
\Omega^{D+2}=F,\quad
\phi(x)=\sqrt{(D+2)(D+3)}M_{pl} \omega(x)
\end{align}
in Eq. \eqref{action:conformaled2}.
The result has
\begin{align}
S=\int d^{4}x d^Dy \sqrt{-\tilde{g}^{4D}}\left[
\frac{M^2_{pl}}{2}\tilde{R}-\frac{1}{2}\tilde{g}^{MN}\partial_M\phi\partial_N\phi-V(\phi)\right].
\end{align}
Under the canonical normalization, we derive the the potential of $\phi$ as
\begin{align}
V(\phi)
&=V_0 \exp\left[\frac{D}{\sqrt{(D+2)(D+3)}}\frac{\phi}{M_{pl}}\right]\left(1-\exp\left[-\sqrt{\frac{D+2}{D+3}}\frac{\phi}{M_{pl}}\right]\right)^2,
\end{align}
where we put $V_0=M^2 M^2_{pl}/8$.
The $\phi$ is regarded as the inflaton.
This potential with $D=0$ is reduced to $V\propto(1-e^{-\sqrt{2/3}\phi/M_{pl}})^2$, which reproduces the original Starobinsky type potential.

\let\doi\relax
\bibliographystyle{utphys28mod}
\bibliography{Ref}

\end{document}